\documentclass[12pt,preprint]{aastex}
\usepackage{spr-astr-addons}
\usepackage{times}
\usepackage{lipsum,amsmath,multicol}
\usepackage{amsmath,amssymb}

\RequirePackage{color}

\begin{document}

\title{Compact stars with quadratic equation of state}
\slugcomment{}
\shorttitle{Short article title}
\shortauthors{Autors et al.}

\author{Sifiso A. Ngubelanga} 
\affil{Astrophysics and Cosmology Research Unit, School of Mathematics, Statistics and Computer Science, University of KwaZulu-Natal, Private Bag X54001, Durban 4000, South Africa
}

\author{Sunil D. Maharaj}
\affil{Astrophysics and Cosmology Research Unit, School of Mathematics, Statistics and Computer Science, University of KwaZulu-Natal, Private Bag X54001, Durban 4000, South Africa.}

\author{Subharthi Ray}
\affil{Astrophysics and Cosmology Research Unit, School of Mathematics, Statistics and Computer Science, University of KwaZulu-Natal, Private Bag X54001, Durban 4000, South Africa.}
\maketitle


\begin{abstract}
We provide new exact solutions to the Einstein-Maxwell system of equations for matter configurations with anisotropy and charge. The spacetime is static and spherically symmetric. A quadratic equation of state is utilised for the matter distribution. By specifying a particular form for one of the gravitational potentials and the electric field intensity we obtain new exact solutions in isotropic coordinates. In our general class of models, an earlier model with a linear equation of state is regained. For particular choices of parameters we regain the masses of the stars PSR J1614-2230, 4U 1608-52, PSR J1903+0327, EXO 1745-248 and SAX J1808.4-3658. A comprehensive physical analysis for the star PSR J1903+0327 reveals that our model is reasonable.
\end{abstract}  

\textit{Keywords}: Exact solutions; anisotropy; electric field intensity; gravitational potential

\section{Introduction}

The modelling of highly dense matter configurations in a general relativistic setting is an important research problem. Recent attempts in this direction include the effects of anisotropy and the electromagnetic field. Some recent results are those of \cite{Mafa}, \cite{Mafa1, Mafa2}, \cite{Maharaj} and \cite{Sunzu, Sunzu1}. However these treatments and others have been completed in the context of Schwarzschild coordinates. There have been fewer investigations involving isotropic coordinates. \cite{Pant} analysed a family of exact solutions of the Einstein-Maxwell field equations in isotropic coordinates. An application to neutron star and quark star with Einstein-Maxwell field equations in isotropic coordinates was performed by \cite{Pradhan}. An investigation of a class of super dense stars models using charged analogues of Hajj-Boutros type relativistic fluid solutions has recently been completed by \cite{Pant1}. In a recent analysis \cite{Ngubelanga} found exact models for a compact stellar object which could be charged and anisotropic with a linear equation of state.  Other
recent investigations that include the effects of anisotropy and the electromagnetic field are contained in the treatments of \cite{x1}, \cite{x2} and \cite{x3}.

A simple generalisation of the linear relation between the energy density and radial pressure is a quadratic equation of state. This allows for more general behaviour in the matter distribution and greater complexity in the model. There is still a debate over the structure of a star as to its composition in terms of nuclear matter, or quark matter, or a hybrid mix of both distributions. It is difficult to find a single equation of state for a matter distribution matching the stellar core (softer quark matter) to the outer regions (stiffer nuclear matter). These issues are highlighted in the treatments of \cite{Cottam}, \cite{Ozel} and \cite{Rodrigues}. A quadratic equation of state which is softer at low densities and stiffer at high densities may be appropriate for describing a hybrid star. This would make it possible to explain the stability of compact stars with masses $\sim2M_\odot$. In a general relativistic context models which are charged and anisotropic were found by \cite{Feroze}. A class of models, generalising the results of \cite{Feroze} and containing models with linear equations of state, was found by \cite{Maharaj1}. These solutions have the desirable property of regularity at the stellar centre. \cite{Mafa2} modelled a charged general relativistic star with a quadratic equation of state. They showed their results were consistent with several masses of stellar objects, in particular with the star PSR J1614-2230. \cite{Malaver} found exact solutions to the field equations for a strange quark model. Analytic and regular models, extending earlier investigations for polytropic distributions, for quark stars and compact objects
with a modified van der Waals equation of state were generated by \cite{x4,x5}.
\cite{Sharma} presented a class of new models using the metric ansatz of \cite{Finch} without assuming any equations of state. Their approach has the remarkable feature of yielding  a quadratic equation of state when appropriate physical bounds are applied. \cite{Thirukkanesh} and \cite{Mafa3} generated exact anisotropic spheres which are uncharged and charged, respectively. These models have a polytropic equation of state in general; however for particular parameter values quadratic equations of state arise.

The aim of this paper is to obtain new exact solutions to the Einstein-Maxwell system of equations. We model charged anisotropic matter distributions in isotropic coordinates by imposing a quadratic equation of state which relates the radial pressure to the energy density. The Einstein-Maxwell field equations in the presence of electric field with anisotropic pressures are presented in section 2. The transformation that has been utilized by \cite{Kustaanheimo}, \cite{Ngubelanga1} and \cite{Ngubelanga} is applied to write the field equations in new equivalent forms. In section 3, we present new classes of exact solutions to the system of equations. We show that the new solution with a quadratic barotropic equation of state contains a known solution by \cite{Ngubelanga} in section 4. In section 5, we regain the masses for the observed objects and study the physical properties of the new exact solutions. We analyse the physical features for the stellar model associated with the star PSR J1903+0327 in section 6. Some concluding remarks are made in section 7.

\section{The model}

We intend to model the interior of a dense star. The line element in isotropic coordinates has the form 

\begin{equation}
\label{eq:g1}  d s^{2} = -A^{2}(r)dt^{2} + B^{2}(r)[dr^{2} + r^{2} (d \theta^{2} + \sin^{2} \theta d \phi^2)],
\end{equation}
in coordinates $(x^a)=(t, r, \theta, \phi)$. The gravitational field is represented by the metric quantities $A(r)$ and $B(r)$ in the metric (\ref{eq:g1}). An anisotropic charged matter distribution has energy momentum of the form 

\begin{eqnarray}
\label{eq:g2}	\nonumber T_{ij} = \mbox{diag}\left(-\rho-\frac{1}{2}E^2, p_{r}-\frac{1}{2}E^2, p_{t}+\frac{1}{2}E^2,  \right. \\
\left. p_{t} +\frac{1}{2}E^2 \right),
\end{eqnarray}
where $\rho$ is the energy density, $p_{r}$ is the radial pressure, $p_{t}$ is the tangential pressure and $E$ is the electric field intensity. A timelike unit four-velocity \textbf{u} where $u^i$ = $\frac{1}{A}\delta^i_{0}$ measures the quantities in equation (\ref{eq:g2}) above.

If we introduce the transformation 

\begin{equation}
\label{eq:g3} x  \equiv  r^{2}, \hspace{0.5cm} L  \equiv  B^{-1},   \hspace{0.5cm}   G \equiv  LA,
\end{equation}

\noindent then the line element can be written in the new form 

\begin{eqnarray}
\label{eq:g4}  \nonumber ds^{2} = -\left(\frac{G}{L}\right)^{2}dt^{2} &+& L^{-2}\left[\frac{1}{4x}dx^{2} \right. \\
&&\left. + x (d \theta^{2} + \sin^{2} \theta d \phi^2)\right],
\end{eqnarray}

\noindent in new variables of $x$. The system of the Einstein-Maxwell field equations can be expressed as 

\begin{eqnarray}
\label{eq:g5} 8\pi \rho + \frac{1}{2} E^{2}   &=& 4[2xLL_{xx}-3(xL_{x}-L)L_{x}],\\
\label{eq:g6}  8\pi p_{r}  -\frac{1}{2} E^{2}   &=& 4L(L-2xL_{x})\frac{G_{x}}{G}\nonumber \\
&& -4(2L-3xL_{x})L_{x},\\
\label{eq:g7} \nonumber  8\pi p_{t}   + \frac{1}{2} E^{2}  &=& 4xL^{2}\frac{G_{xx}}{G}+4L(L-2xL_{x})\frac{G_{x}}{G} \\
&&-4(2L-3xL_{x})L_{x}-8xLL_{xx},\\
\label{eq:g8}  \sigma^2 &=& \frac{1}{4\pi x} L^2  (E+xE_{x})^2,
\end{eqnarray}
in terms of new variables by utilizing transformation (\ref{eq:g3}). The subscript $``x"$ denotes 
a derivative with respect to the new variable $x$. In terms of new variables in (\ref{eq:g3}) the condition 
of pressure anisotropy has the form

\begin{equation}
\label{eq:g9}		 \frac{G_{xx}}{G}-2\frac{L_{xx}}{L} = \frac{(8\pi \Delta + E^2)}{4xL^2},
\end{equation}
where the quantity $\Delta=p_{t}-p_{r}$ is the measure of anisotropy. The mass function 
has the form  

\begin{equation}
\label{eq:g10}	m(x) = 2\pi  \int^x_{0}{\left[\sqrt{\omega} \rho(\omega) +\frac{E^2}{8\pi}\right]d\omega},
\end{equation}
in new coordinates. The mass function represents the mass within the radius $x$ of the sphere.

We assume the quadratic equation of state of the form

\begin{equation}
\label{eq:g11} 	p_{r} = \eta \rho^2+\alpha \rho - \beta,
\end{equation}
relating the radial pressure $p_{r}$ to the energy density $\rho$, and where $\eta$, $\alpha$ and 
$\beta$ are arbitrary constants. This is a simple generalisation of a linear equation of state which is regained when $\eta=0$. With the inclusion of the quadratic equation of state, the Einstein-Maxwell 
system of equations (\ref{eq:g5})-(\ref{eq:g8}) with the charged anisotropic fluid spheres can be expressed 
as

\begin{eqnarray}
\label{eq:g12}		8\pi \rho   	 &=&	 4[2xLL_{xx}-3(xL_{x}-L)L_{x}]- \frac{1}{2} E^{2},\\
\label{eq:g13}	 	      p_{r}	 &=&	 \eta \rho^2+\alpha \rho - \beta,\\
\label{eq:g14}		      p_{t}	 &=& 	   p_{r}  +    \Delta,\\
\label{eq:g15} 	 8\pi     \Delta 	 &=&	 4xL^2\frac{G_{xx}}{G}+4L(L-2xL_{x})\frac{G_{x}}{G}\nonumber \\
&& -\frac{\eta}{32\pi}\left[16xLL_{xx}-24(xL_{x}-L)L_{x}-E^2 \right]^2\nonumber\\		\nonumber\\
					& &	-8(1+\alpha)xLL_{xx}+12(1+\alpha)xL^2_{x}\nonumber\\		\nonumber\\		
					& &    -4(2+3\alpha)LL_{x}   -\frac{(1-\alpha)E^2}{2}+8 \pi \beta,\\														 \nonumber\\
\label{eq:g16}	\frac{G_{x}}{G}&=&   \frac{\eta\left[16xLL_{xx}-24(xL_{x}-L)L_{x}-E^2\right]^2}{128\pi L(L-2xL_{x})}  \nonumber\\	 \nonumber\\
					& & +\frac{2\alpha x  L_{xx}}{(L-2xL_{x})} -\frac{3(1+\alpha)  x L^2_{x}}{L(L-2xL_{x})}\nonumber \\
					&& +\frac{(2+3\alpha) L_{x}}{(L-2xL_{x})}-\frac{(1+\alpha)E^2}{8 L(L-2xL_{x})}\nonumber\\			 \nonumber\\
					& &-\frac{2\pi \beta}{L(L-2xL_{x})},\\			 \nonumber\\
\label{eq:g17}	 \sigma^2	 &=& 	\frac{1}{4\pi x} L^2  (E+x E_{x})^2.
\end{eqnarray}

It is crucial to note the non-linearity in both the functions $L$ and $G$ in the system 
(\ref{eq:g12})-(\ref{eq:g17}) which is increased because of the appearance of terms containing the parameter $\eta$. This system of equations contains six variables involving the matter and the electromagnetic quantities ($\rho$, $p_{r}$, $p_{t}$, $\Delta$, $E$ and $\sigma$) and two gravitational 
potentials ($L$ and $G$). It should also be highlighted that there are only six independent equations in this 
system of equations. Integration of such systems is not easy to perform due to nonlinearity and the 
fact that there are more unknown functions than the independent field equations. In order to integrate and 
obtain some exact solutions, the above mentioned facts suggest that we need to choose the form for two of 
the quantities mentioned above. The system of equations (\ref{eq:g12})-(\ref{eq:g17}) is similar to the
field equations of \cite{Ngubelanga}; however in our case the equation of state is
quadratic. In their treatment they utilized the linear equation of state, $i.e.,$
$\eta=0$ so that $p_{r}=\alpha \rho-\beta$.

The interior metric (\ref{eq:g1}) with the charged matter distribution should match the exterior 
spacetime which is given by 
\begin{eqnarray}
 \label{eq:g18}	ds^2 &=& -\left(	1-\frac{2M}{R}+\frac{q^2}{R^2}\right)dt^2+\left(1-\frac{2M}{R} \right. \nonumber\\	\nonumber\\	
				& &\left. +\frac{q^2}{R^2}\right)^{-1}dR^2 +R^2(d\theta^{2} + \sin^{2}\theta d\phi^2),
\end{eqnarray}

\noindent in coordinates $(x^a) = (t, R, \theta, \phi)$. In (\ref{eq:g18}) the total mass and the total charge of the sphere are denoted by $M$ and $q^2$, respectively. The exterior spacetime (\ref{eq:g18}) is referred to as the Reissner-Nordstr\"om metric. The junction conditions at the stellar surface are obtained by matching the first and the second fundamental forms for the interior metric (\ref{eq:g1}) and the exterior metric (\ref{eq:g18}). The conditions are as follows 

\begin{eqnarray}
\label{eq:g19} 	A_{s} &=& \left(1-\frac{2M}{R}+\frac{q^2}{R^2}\right)^{\frac{1}{2}},\\	\nonumber\\
\label{eq:g20}	R_{s} &=& r_{s} B_{s},\\		\nonumber\\
\label{eq:g21}	\left(\frac{B'}{B}+\frac{1}{r}\right)_{s}r_{s}&=&  \left(1-\frac{2M}{R}+\frac{q^2}{R^2}\right)^{\frac{1}{2}},\\	\nonumber\\
\label{eq:g22}	r_{s}(A')_{s} &=& \frac{M}{R}-\frac{q^2}{R^2},
\end{eqnarray}

\noindent evaluated at the boundary of the star $r=s$. In isotropic coordinates the boundary conditions are given by equations 
(\ref{eq:g19})-(\ref{eq:g22}).

\section{Exact models}

Our purpose is to generate new exact solutions to the Einstein-Maxwell system of equations 
(\ref{eq:g12})-(\ref{eq:g17}). The integration is achieved by choosing physical reasonable 
forms for the electric field $E$ and the gravitational potential $L$. We make the particular choice 

\begin{eqnarray}
\label{eq:g23}	L 	&=& a+bx,\\
\label{eq:g24}	E^2	&=& x(c+dx),
\end{eqnarray}
where $a$, $b$, $c$ and $d$ are real constants. The potential $L$ and the electric 
field intensity $E^2$, respectively, are selected to be a linear function and a quadratic function in the 
variable $x$. Similar choices for $L$ and $E$ were made by \cite{Ngubelanga} for a linear equation of state leading to acceptable stellar configurations; we expect this to also carry through with the addition of a quadratic term in the equation of state. On applying (\ref{eq:g23}) and (\ref{eq:g24}), equation (\ref{eq:g16}) becomes 

\begin{eqnarray}
\label{eq:g25}		\frac{G_{x}}{G}&=&\frac{\eta\left[24ab-(c+dx)x\right]^2}{128 \pi (a-bx)(a+bx)}+\frac{b(2+3\alpha)}{(a-bx)}\nonumber\\		\nonumber\\
					& & -\frac{[16\pi \beta+(1+\alpha)(24b^2+c+dx)x]}{8(a-bx)(a+bx)},	
\end{eqnarray}

\noindent  which is a first order equation in potential $G$. We integrate (\ref{eq:g25}) to obtain 

\begin{equation}
\label{eq:g26} G(x) =   K(a-bx)^{\Psi}(a+bx)^{\Phi} e^{N(x)},
\end{equation}
where $K$ is the constant of integration. The function $N(x)$ and the constants 
$\Psi$ and $\Phi$ are given explicitly by

\begin{eqnarray}
\label{eq:g27}		  N(x)	&=& \frac{x}{384 \pi b^4}\left\{48\pi b^2d(1+\alpha)\right.\nonumber\\ 	\nonumber\\ 
				& & \left. -3 \eta \left[b^2 c^2+ad(ad-48b^3)\right] \right. \nonumber \\
				&& \left. -b^2d\eta(3c+dx)x\right\},\\	\nonumber\\
\label{eq:g28}		\Psi	&=&\frac{1}{256\pi ab^5 }\left\{16\pi ab^2(1+\alpha)(bc+ad)\right.\nonumber\\		\nonumber\\
				&& \left. -128\pi b^4\left[ab(1+3\alpha)-2\pi \beta \right] \right. \nonumber \\
					& &\left. -a^2 \eta\left[a^2d^2+b(c-24b^2)\right.\right.\nonumber \\
					&& \left.\left. \times (2ad+bc-24b^3)\right]\right\},\\	\nonumber\\
\label{eq:g29}		\Phi	&=&\frac{1}{256\pi ab^5}\left\{16\pi ab^2(1+\alpha)\left[b(24b^2+c)-ad\right]\right.\nonumber\\		\nonumber\\
				& & \left. -256 \pi^2 b^4 \beta +a^2\eta \left[a^2d^2\right. \right.\nonumber \\
				&& \left. \left. -b(24b^2+c)(2ad-bc-24b^3)\right] \right\},
\end{eqnarray}
where the constants $a\neq0$ and $b\neq0$ to avoid singularity. An exact solution can then be 
found to the Einstein-Maxwell system. The metric (\ref{eq:g1}) has the form 

\begin{eqnarray}
\label{eq:g30}	ds^2 &=& - K (a-br^2)^{2\Psi}(a+br^2)^{2(\Phi -1)}e^{2N(r)}dt^2\nonumber\\
				& & + (a+br^2)^{-2}[dr^2+r^2(d\theta^2+\sin^2\theta d\phi^2)],
\end{eqnarray}	
where $K$ is the constant of integration. The function $N(r)$ and the constants $\Psi$ and 
$\Phi$ are given explicitly by (\ref{eq:g27})-(\ref{eq:g29}).

Since equation   (\ref{eq:g25}) has been integrated, then we can generate an
exact model for the system of equations  (\ref{eq:g12})-(\ref{eq:g17})  in terms of the radial coordinate ``$r$" which has the form

\begin{eqnarray}
\label{eq:g31}	     8\pi	\rho       &=& 12ab-\left(\frac{1}{2}c+\frac{1}{2}dr^2\right)r^2,\\	
\label{eq:g32}		p_{r}    &=&    \eta \rho^2+\alpha \rho-\beta, \\	
\label{eq:g33}	 	p_{t}    &=& p_{r}+\Delta,
\end{eqnarray}

\begin{eqnarray}
\label{eq:g34}	8\pi       \Delta  &=& 	\frac{4b\Psi(a+br^2)r^2}{(a-br^2)^2}\left[b(\Psi-1)(a+br^2)\right. \nonumber \\
&& \left. -2(a-br^2)\left(b\Phi +N'(a+br^2)\right)\right]	\nonumber\\
		& & -\frac{\eta}{32\pi}\left[24ab-r^2(c+dr^2)\right]^2\nonumber \\
		&& +4(a+br^2)\left[a-br^2(1-2\Phi)\right]N'	\nonumber\\	
		& & +4r^2(a+br^2)^2(N'^2+N'') \nonumber \\
		&& +4b\left[\Phi(a-br^2)-(a+br^2)(2+3\alpha+\Psi)\right]		\nonumber\\
		& & +4b^2\left[3(1+\alpha)+\Phi(\Phi-1)\right]r^2\nonumber \\
		&& -\frac{1}{2}(1-\alpha)(c+dr^2)r^2+8\pi \beta,\\	\nonumber\\
\label{eq:g35}	  \sigma^2      &=& \frac{(a+br^2)^2}{16 \pi^2 (c+dr^2)}\left[3c+4dr^2\right]^2,	\\	\nonumber\\
\label{eq:g36}		E^2    &=& r^2(c+dr^2).
\end{eqnarray}
It is interesting to note that our model is of a simple form and all physical quantities are expressed in terms of elementary functions where the function $N(r)$ and the constants $\Psi$ and $\Phi$ are given in (\ref{eq:g27})-(\ref{eq:g29}), respectively. For this model the mass function is given by

\begin{equation}
\label{eq:g37}	m(r)     = \frac{1}{2} r^3 \left[\frac{(12ab+c)}{3}-\frac{(c-2d)r^2}{10}-\frac{dr^4}{14}\right].
\end{equation}

A charged anisotropic star with quadratic equation of state may be modeled by the above solution (\ref{eq:g31})-(\ref{eq:g36}).

\section{The linear case}

When $\eta=0$ then the equation of state becomes 

\begin{equation}
\label{eq:g38} 		p_{r} = \alpha \rho - \beta, 
\end{equation}
which is linear. We observe that the case (\ref{eq:g38}) reduces to the \cite{Ngubelanga} model. Our result is a generalisation with a quadratic equation of state. All the results in \cite{Ngubelanga} can be regained as a special case from the exact solution (\ref{eq:g31})-(\ref{eq:g36}). The relationship (\ref{eq:g38}) is consistent with the stars PSR J1614-2230, Vela X-1, PSR J1903+0327, 4U 1820-30 and SAX J1808.4-3658 as demonstrated in their analysis. In particular for the parameter values $a=1.96819$, $b=0.5$, $c=0.01$, $d=0.01$ and $\alpha=0.931$ we can produce the mass $1.97 M_\odot$. This stellar mass corresponds to the astronomical object PSR J1614-2230.

\section{The quadratic case}

From the exact solution (\ref{eq:g31})-(\ref{eq:g36}) we observe that the quantities associated with the matter field and the electromagnetic field are well behaved. The electric field vanishes at the stellar centre $r=0$. The matter density $\rho$ and the proper charge density $\sigma$ remain finite at the centre. At the centre of the star we can write 

\begin{eqnarray}
\label{eq:g39}	\rho_{0} & = & \frac{3ab}{2\pi},\\
\label{eq:g40}	  p_{r0} & = & \eta \left(\frac{3ab}{2\pi}\right)^2+\alpha \left(\frac{3ab}{2\pi}\right)-\beta,
\end{eqnarray}
which are finite values. For the electromagnetic quantities we have 

\begin{eqnarray}
\label{eq:g41}		\sigma^2_{0} & = & \frac{1}{c} \left(\frac{3ac}{4\pi}^2\right), \\
\label{eq:g42} 		E^2_{0}      & = & 0,
\end{eqnarray}

\noindent which are nonsingular at the centre. For the pressure anisotropy we have 

\begin{eqnarray}
\label{eq:g43}		\Delta_{0} = -\frac{3ab\eta}{32\pi^{2}} &+& \frac{a^2}{2\pi}N'(0)\nonumber\\
&+&\frac{ab}{2\pi}[\Phi -(2+3 \alpha + \Psi)] + \beta,	
\end{eqnarray}

\noindent at $r=0$. The metric functions $A$ and $B$ are regular at $r=0$. Therefore all physical and gravitational quantities are well behaved in the core regions of the star. For the star to remain stable it is required that $\Delta=0$ at $r=0$; we demonstrate that this happens in the next section using a graphical treatment. The mass remains finite and also depends on the parameters $c$ and $d$ which are associated with charge. 

The speed of sound is defined by 

\begin{equation}
\label{eq:g44}		v^2 = \frac{dp_{r}}{d\rho},
\end{equation}
where we must have $v<1$ to maintain causality. Also we must have zero  radial pressure at the boundary for a stable configuration of the compact object. This will ensure consistency of the matching conditions (\ref{eq:g19})-(\ref{eq:g22}) at the surface and continuity of the metrics (\ref{eq:g1}) and (\ref{eq:g18}) at the surface. For a finite value of the density at the surface along with the zero pressure, we require 

\begin{equation}
\label{eq:45}		\rho_{s} = \frac{3ab}{2\pi} -\frac{1}{16\pi}(c+d),
\end{equation}
in geometric units by fixing the radius of the star at $r=1$. Then (\ref{eq:g32}) restricts the parameter $\beta$ by 

\begin{equation}
\label{eq:g46} 		\beta_{s} = \eta \rho^2_{s}+\alpha \rho_{s}.
\end{equation}

When $\eta=0$ then (\ref{eq:g46}) reduces to the corresponding expression of \cite{Ngubelanga}. In our subsequent analysis throughout we choose the parameter values $\alpha=0.931$ and $\eta=3.185$ since they produce relativistic compact stars with desirable physical features.

It is to be noted that the relation for the mass in (\ref{eq:g37}) is free from the equation of state parameters $\eta$, $\alpha$ and $\beta$. Hence the value of the mass will be indistinguishable from that of the results in the linear case as per \cite{Ngubelanga} if we select the same parameter values. To tackle this issue, we have matched the values of the gravitational potentials ($A^2$) at the stellar surface ($r=1$) for both the linear and the quadratic cases, showing that for an exterior observer, the gravitational potential should be the same. Thus the effect of the parameters in the quadratic equation of state comes into the system through the $\Phi$ and the $\Psi$  terms in the gravitational potentials.

It is possible to give numerical values to quantities in our exact solutions. We have considered values for five compact objects for which reliable data exists. The objects selected are PSR J1614-2230 studied by \cite{Demorest}, 4U 1608-52 investigated by \cite{Guver}, PSR J1903+0327 analysed by \cite{Freire}, EXO 1745-248 studied by \cite{Ozel1} and SAX J1808.4-3658 considered by \cite{Elebert}. We vary the parameter $a$ in (\ref{eq:g37}) and assign fixed values for $b=0.504167$, $c=0.01$ and $d=0.01$. This permits us to generate numerical values for the stellar masses for the five astronomical objects listed in Table \ref{tab:5stars1}. We have used small values for the parameters $c$ and $d$ which introduce charge into the system to ensure that the electromagnetic contribution is small. We find that the observed masses vary between $0.9 M_\odot$ to $1.97 M_\odot$. Values for the central density $\rho_{0}$, central radial pressure $p_{r0}$ and  surface density $\rho_{s}$ lie in the expected range.

\begin{table*}[h]
\caption{Mass $m$, central density $\rho_{0}$, central radial pressure $p_{r0}$ and surface density $\rho_{s}$ of different stars corresponding to the parameters $b=0.504167$, $c=0.01$, $d=0.01$, $\alpha=0.931$ and $\eta=3.185$}
\begin{center}
\begin{tabular}{|c|c|c|c|c|c|c|}
\hline
Star		&Observed mass $m$ & $a$ & $\rho_{0}$ & $p_{r0}$  & $\rho_{s}$  \\
\hline
	PSR  J1614-2230			& 1.97  & 1.95192  & 0.469871 & 0.00156084  & 0.469473\\
	4U 1608-52				& 1.74  & 1.72382  & 0.414962 & 0.00142167  & 0.414565\\
	PSR J1903+0327			& 1.667 & 1.65143  & 0.397535 & 0.0013775   & 0.397137\\
	EXO 1745-248			& 1.3   & 1.28746  & 0.30992  & 0.00115543  & 0.309522\\
	SAX J1808.4-3658		& 0.9   & 0.890767 & 0.214427 & 0.000913404 & 0.214029\\
\hline
\end{tabular}
\end{center}
\label{tab:5stars1}
\end{table*}

\section{The star PSR J1903+0327 }

The parameter value $a=1.65143$ generates the mass 1.667 $M_\odot$ which corresponds to the star PSR J1903+0327. We use this parameter value for $a$ to analyse the variation of the physical features associated with the matter, charge and gravity field within the star. 

Table \ref{tab:anis1} represents the variation of density $\rho$, radial pressure $p_{r}$, tangential pressure $p_{t}$ and anisotropy $\Delta$ within the star. The quantities $\rho$ and $p_{r}$ are decreasing functions. The radial pressure $p_{r}$ vanishes at $r=1$ determining the boundary which is the requirement for a compact star. The tangential pressure $p_{t}$ has finite values. The anisotropy $\Delta$ remains finite and has the value $\Delta=0$ at $r=0$ which is required for stability. Table \ref{tab:mass1} presents the behaviour of the mass $m$, electric field $E$ and charge density $\sigma$. The mass increases as $r$ grows larger. The electric field $E$ and charge density $\sigma$ are finite and nonsingular throughout the star with $E=0$ at $r=0$. The effect of the charge is incorporated through the parameters $c$ and $d$. Tables \ref{tab:chrgd1} and \ref{tab:chrgc1} represent the total charge in the star with $r=1$ fixed at the stellar surface. It is clear that the parameter $d$ has a greater effect than that of the parameter $c$ which makes the star more charged. The metric functions $A^2$ and $B^2$ are evaluated in Table \ref{tab:pot1} for the set of parameter values corresponding to PSR J1903+0327 through the interior of the star. The values obtained for the metric functions indicate that the potentials are regular and positive.

\begin{figure}
\begin{center}
\includegraphics[width=.45\textwidth]{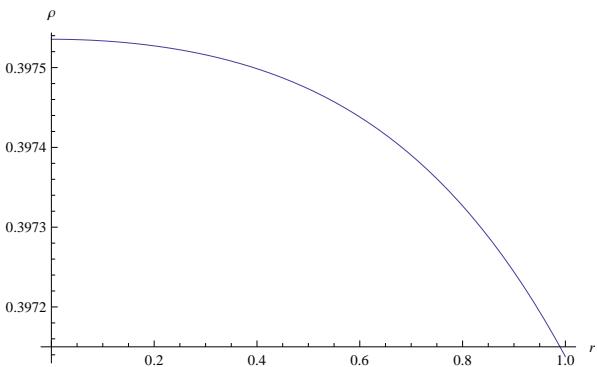}
\caption{Variation of density ($y$-axis) with the radius}
\label{fig:rho1}
\end{center}
\end{figure}

\begin{figure}
\begin{center}
\includegraphics[width=.45\textwidth]{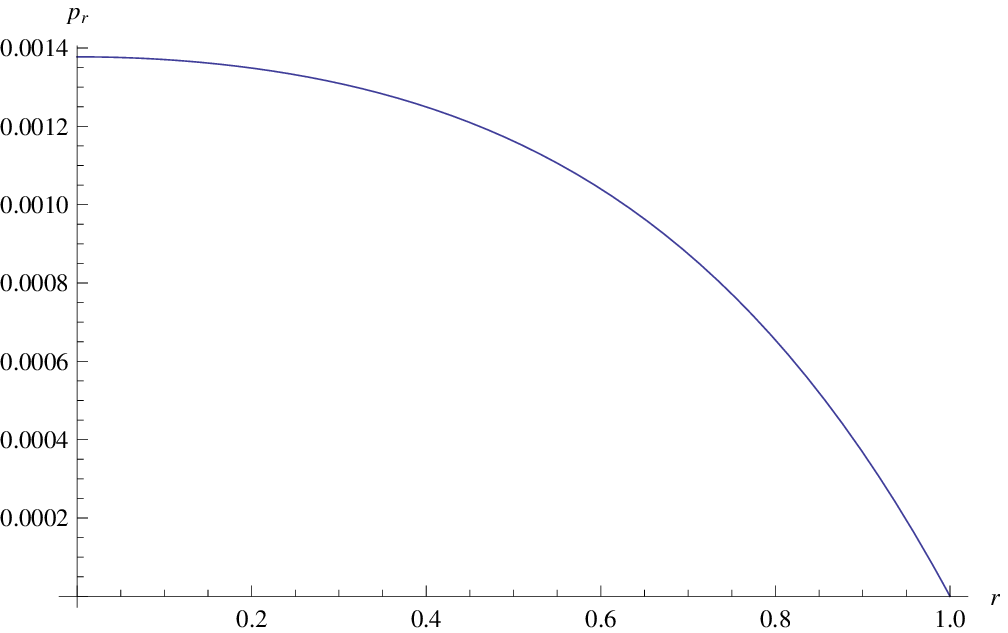}
\caption{Variation of pressure ($y$-axis) with the radius}
\label{fig:pr1}
\end{center}
\end{figure}

\begin{figure}
\begin{center}
\includegraphics[width=.45\textwidth]{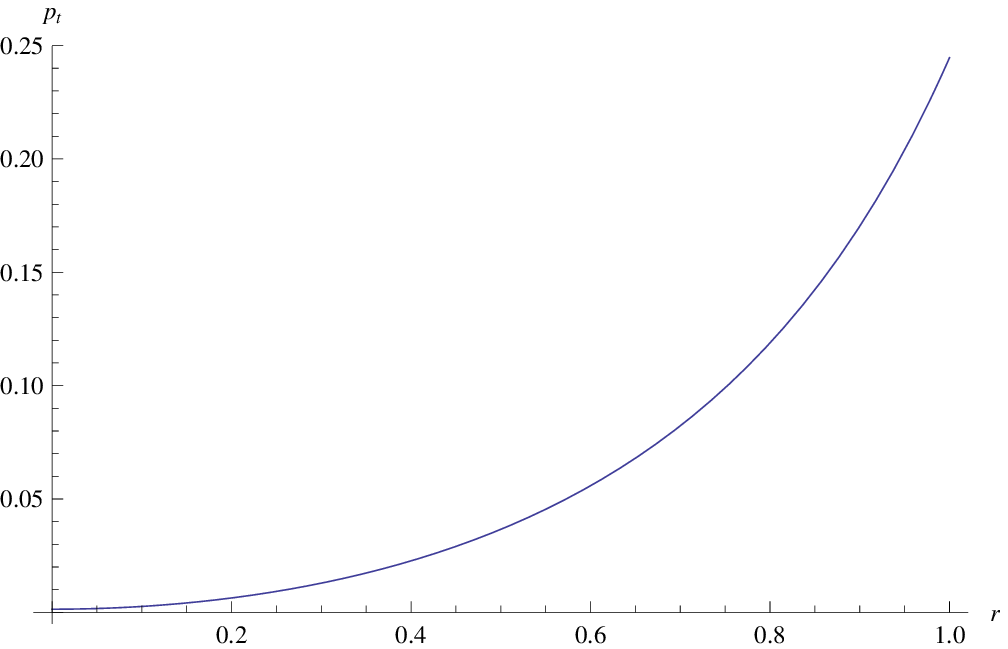}
\caption{Variation of tangential pressure ($y$-axis) with the radius.}
\label{fig:pt1}
\end{center}
\end{figure}

\begin{figure}
\begin{center}
\includegraphics[width=.45\textwidth]{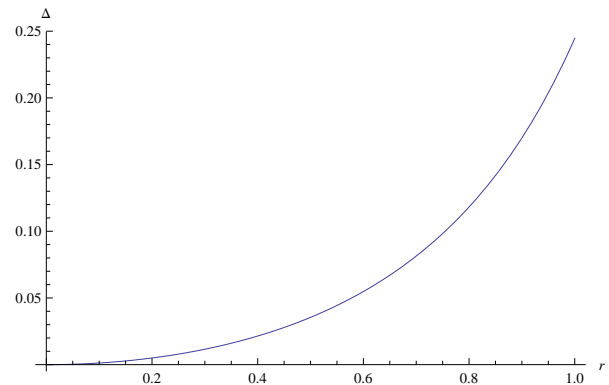}
\caption{Variation of measure of anisotropy ($y$-axis) with the radius.}
\label{fig:delta1}
\end{center}
\end{figure}

\begin{figure}
\begin{center}
\includegraphics[width=.45\textwidth]{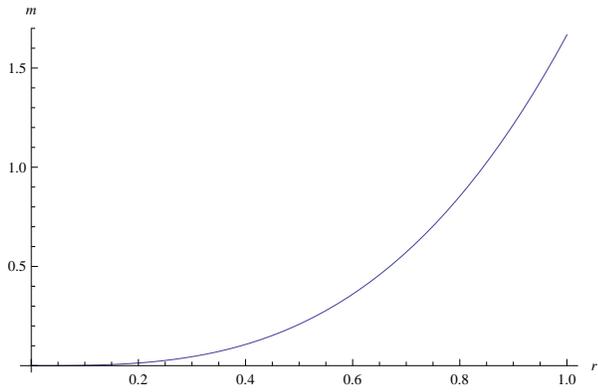}
\caption{Variation of mass ($y$-axis) with the radius.}
\label{fig:mass1}
\end{center}
\end{figure}

\begin{figure}
\begin{center}
\includegraphics[width=.45\textwidth]{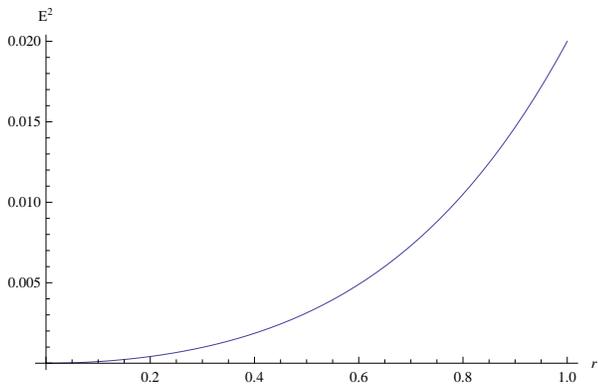}
\caption{Square of Electric field intensity with the radius.}
\label{fig:elect1}
\end{center}
\end{figure}

\begin{figure}
\begin{center}
\includegraphics[width=.45\textwidth]{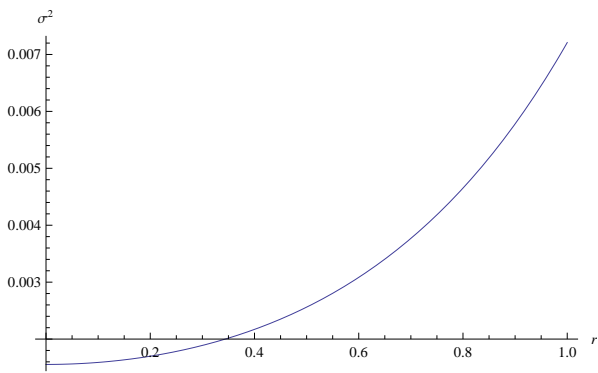}
\caption{Charge density against the radius.}
\label{fig:chrg1}
\end{center}
\end{figure}

\begin{figure}
\begin{center}
\includegraphics[width=.45\textwidth]{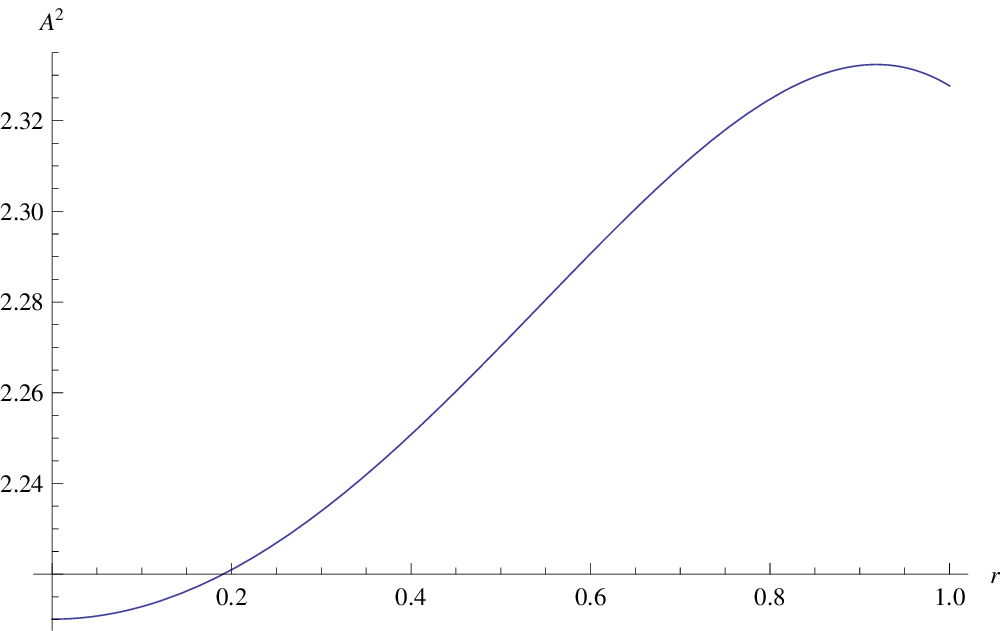}
\caption{Variation of the gravitational potential $A^2$ ($y$-axis) with the radius.}
\label{fig:potA1}
\end{center}
\end{figure}

\begin{figure}
\begin{center}
\includegraphics[width=.45\textwidth]{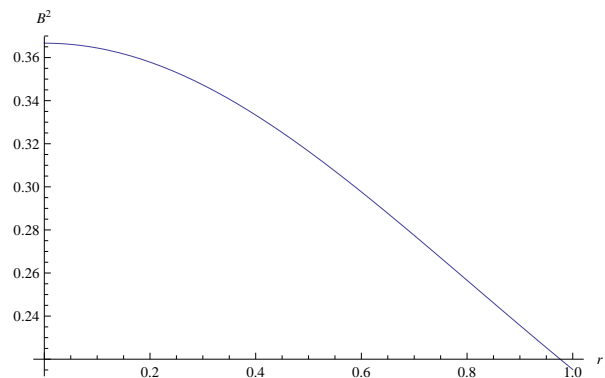}
\caption{Variation of the gravitational potential $B^2$ ($y$-axis) with the radius.}
\label{fig:potB1}
\end{center}
\end{figure}

A graphical analysis provides deeper insight into the behaviour of the physical features. We have presented plots for the density (Fig. \ref{fig:rho1}), radial pressure (Fig. \ref{fig:pr1}), tangential pressure (Fig. \ref{fig:pt1}), pressure anisotropy (Fig. \ref{fig:delta1}), mass (Fig. \ref{fig:mass1}), electric field intensity (Fig. \ref{fig:elect1}), charge density (Fig. \ref{fig:chrg1}) and metric functions (Fig. \ref{fig:potA1} and Fig. \ref{fig:potB1}). It is clear that all the quantities have regular profiles from the various plots that have been generated. \cite{Ngubelanga} using a linear equation of state also studied particular observed stars in general relativity. Our results in this paper with a quadratic equation of state are broadly consistent with their results.

\begin{table*}[h]
\caption{Variation of energy density $\rho$, radial pressure $p_{r}$, tangential pressure $p_{t}$ and measure of anisotropy $\Delta$ from the centre to the surface with parameters $a=1.65143$, $b=0.504167$, $c=0.01$, $d=0.01$, $\alpha=0.931$ and $\eta=3.185$}
\begin{center}
\begin{tabular}{|c|c|c|c|c|c|c|c|c|c|}
\hline
$r$ 		& $\rho$ & $p_{r}$ & $p_{t}$ & $\Delta$    \\
\hline

	0   & 0.397536 & 0.0013775   & 0.0013775  & 0\\
	0.1 & 0.397534 & 0.00137054  & 0.00259535 & 0.00122481\\
	0.2 & 0.397527 & 0.00134884  & 0.00633559 & 0.00498676\\
	0.3 & 0.397516 & 0.00130991  & 0.0128702  & 0.0115603\\
	0.4 & 0.397499 & 0.00124963  & 0.0226954  & 0.0214458\\
	0.5 & 0.397473 & 0.0011622   & 0.0366066  & 0.0354444\\
	0.6 & 0.397438 & 0.00104019  & 0.0558307  & 0.0547905\\
	0.7 & 0.39739  & 0.000874527 & 0.0822543  & 0.0813798\\
	0.8 & 0.397327 & 0.000654462 & 0.118827   & 0.118173\\
	0.9 & 0.397244 & 0.000367625 & 0.170302   & 0.169934\\
	1   & 0.397138 & 0 			 & 0.244665   & 0.244665\\
\hline
\end{tabular}
\end{center}
\label{tab:anis1}
\end{table*}

\begin{table*}[h]
\caption{Variation of mass $m$, electric field intensity $E^2$ and charge density $\sigma^2$ for charged bodies from centre to the surface with parameters $a=1.65143$, $b=0.504167$, $c=0.01$ and $d=0.01$}
\begin{center}
\begin{tabular}{|c|c|c|c|}
\hline
$r$ 		 & $m$ & $E^2$ & $\sigma^2$   \\
\hline
	0		& 0    		 & 0 		& 0.00155433\\
	0.1		& 0.00166686 & 0.000101 & 0.00158992\\
	0.2 	& 0.013335   & 0.000416 & 0.00169896\\
	0.3		& 0.0450063  & 0.000981 & 0.00188841\\
	0.4		& 0.106684   & 0.001856 & 0.00217005\\
	0.5 	& 0.20837    & 0.003125 & 0.00256092\\
	0.6		& 0.360071   & 0.004896 & 0.00308389\\
	0.7		& 0.571787   & 0.007301 & 0.00376848\\
	0.8		& 0.853521   & 0.010496 & 0.00465182\\
	0.9		& 1.21527    & 0.014661 & 0.00577994\\
	1		& 1.667      & 0.02     & 0.00720911\\
\hline
\end{tabular}
\end{center}
\label{tab:mass1}
\end{table*}

\begin{table*}[h]
\caption{Electric field intensity $E^2$ and charge density $\sigma^2$ $(r=1)$ with parameters $r=1$, $a=1.65143$, $b=0.504167$ and $c=0.01$.}
\begin{center}
\begin{tabular}{|c|c|c|}
\hline
$d$ 		& $\sigma^2$ & $E^2$\\
\hline
	0			& 0.00264824 & 0.01\\
	0.1			& 0.0494606  & 0.11\\
	0.2 		& 0.0965278  & 0.21\\
	0.3			& 0.143603   & 0.31\\
	0.4			& 0.190681   & 0.41\\
	0.5 		& 0.237759   & 0.51\\
	0.6		    & 0.284838   & 0.61\\
	0.7 		& 0.331917   & 0.71\\
	0.8 		& 0.378997   & 0.81\\
	0.9 		& 0.426076   & 0.91\\
	1      		& 0.473156   & 1.01\\
\hline
\end{tabular}
\end{center}
\label{tab:chrgd1}
\end{table*}

\begin{table*}[h]
\caption{Electric field intensity $E^2$ and charge density $\sigma^2$ with parameters $r=1$, $a=1.65143$, $b=0.504167$ and $d=0.01$.}
\begin{center}
\begin{tabular}{|c|c|c|}
\hline
$c$ 	& $\sigma^2$ & $E^2$\\
\hline
	0		&  0.00470799 & 0.01 \\
	0.1		&  0.0309229  & 0.11 \\
	0.2		&  0.0573926  & 0.21 \\
	0.3		&  0.0838705  & 0.31 \\
	0.4		&  0.110351   & 0.41 \\
	0.5		&  0.136832   & 0.51 \\
	0.6		&  0.163313   & 0.61 \\
	0.7		&  0.189795   & 0.71 \\
	0.8		&  0.216277   & 0.81 \\
	0.9		&  0.242759   & 0.91 \\
	1		&  0.269241   & 1.01 \\
\hline
\end{tabular}
\end{center}
\label{tab:chrgc1}
\end{table*}

\begin{table*}[h]
\caption{Potentials $A^2$ and $B^2$ with varying radius with parameters $a=1.65143$, $b=0.504167$, $c=0.01$, $d=0.01$, $\alpha=0.931$ and $\eta=3.185$.}
\begin{center}
\begin{tabular}{|c|c|c|}
\hline
$r$		& $A^2$ & $B^2$\\
\hline

 	0 	& 2.21008 	& 0.366674\\
	0.1	& 2.21284	& 0.364445\\
	0.2 & 2.22096 	& 0.35788\\
	0.3 & 2.23391	& 0.347325\\
	0.4 & 2.25077 	& 0.333316\\
	0.5 & 2.27027 	& 0.316515\\
	0.6 & 2.29067 	& 0.297652\\
	0.7 & 2.30976 	& 0.277454\\
	0.8 & 2.32474 	& 0.256604\\
	0.9 & 2.33215 	& 0.235694\\
	1 	& 2.32766 	& 0.215211\\
\hline
\end{tabular}
\end{center}
\label{tab:pot1}
\end{table*}

\section{Discussion}

Our objective in this paper was to find new exact solutions to the Einstein-Maxwell field equations for matter configurations with anisotropy and charge in isotropic coordinates. We selected the barotropic equation of state to be quadratic which relates the radial pressure $p_{r}$ to the energy density $\rho$. The classes of exact solutions (\ref{eq:g31})-(\ref{eq:g36}) to the Einstein-Maxwell field equations were shown to be physically acceptable. The tables for charge and matter variables suggest that they represent physically reasonable configurations. By choosing to fix the parameters $b=0.504167$, $c=0.01$, $d=0.01$, $\alpha=0.931$ and $\eta=3.185$ and varying the parameter $a$ in Table \ref{tab:5stars1}, we regained the masses for the stellar objects PSR J1614-2230, 4U 1608-52, PSR J1903+0327, EXO 1745-248 and SAX J1808.4-3658. We fixed the parameter $a$ and used the star PSR J1903+0327 which has the mass 1.667 $M_\odot$, to produce tables and graphical plots for relevant quantities related to the metric, matter and charge. We made the particular choices $a=1.65143$, $b=0.504167$, $c=0.01$, $d=0.01$, $\alpha=0.931$ and $\eta=3.185$ to perform graphical plots using the software package Mathematica. Our graphical approach suggests that the model for the star PSR J1903+0327 is well behaved. The introduction of the quadratic parameter $\eta$ in the equation of state $p_{r}=\eta \rho^2+\alpha \rho-\beta$ does produce a new exact solution of the Einstein-Maxwell system which is qualitatively different from the linear case $p_{r}=\alpha \rho -\beta$. However the quadratic equation of state shall produce models which can be related to observed stellar objects.

\end{document}